\documentclass[]{emulateapj}
\usepackage{natbib}
\usepackage{psfig}

\begin{document}
\title {The nature of the DLS fast transients}
\author{S. R. Kulkarni and A. Rau}
\affil {Caltech Optical Observatories 105-24,\\
California Institute of Technology, Pasadena, CA 91125, USA}

\begin {abstract} 

The discovery and study  of highly transient sources, especially those
which rise to  high brightness and then fade to  obscurity, has been a
major  part  of  modern  astrophysics.  Well  known  examples  include
supernovae and  novae.  A  byproduct of the  Deep Lens Survey  was the
discovery of three transients which varied on a timescale of less than
an hour.  All  three had faint and red  counterparts, the brightest of
which  was identified  with an  M  star.  However,  the remaining  two
showed hints of an extragalactic origin, one had a spatially extended
counterpart and the other appeared in projection on the outskirts of a
bright  elliptical galaxy.   If these  two sources  were really  of an
extragalactic  origin then  the two  events represent  a new  class of
exotic explosive transients.   We undertook spectroscopic observations
with the  Keck telescope and find  the two counterparts  are also late
type Galactic dwarfs. Our main conclusion is that flares from M dwarfs
constitute  a dense  foreground fog  and dominate  over  any plausible
class  of extragalactic  fast transients  by  at least  two orders  of
magnitude.  Overcoming this fog  will likely require dedicated surveys
with  careful optimization  of  target field  location, filter(s)  and
cadence, pre-search imaging to filter  out late type dwarfs and a well
planned rapid followup plan.

\end {abstract}

\section{Background}

In the first half of the twentieth century the search for and study
of variable sources was a major focus of optical astronomy. 
This area is 
widely expected to undergo a renaissance, thanks
to significant technological changes (sensors, computing, storage
and dissemination). Apart from existing wide-field
imagers dedicated  facilities such as the Panoramic
Survey Telescope \&\ Rapid Response System (Pan-STARRS; \citealt{kab+02})
and the Large Synoptic Survey Telescope (LSST) are motivated in
part to explore transient phenomena in the optical sky. 

An excellent precursor of  the planned optical searches for transients
is  the   Deep  Lens  Survey\footnote{http://dls.physics.ucdavis.edu/}
(DLS). This survey  is notable for its comprehensive  scope and almost
real  time alerts \citep{bwb+04},  The Deep  Lens Survey  netted three
{\em  fast}\footnotemark\footnotetext{Following   the  terminology  of
X-ray astronomy we  refer to transients which change  on timescales of
an hour or less as fast transients.} transients.  The quiescent of the
brightest transient was  identified with an M dwarf  and the transient
itself attributed to  a flare from the M dwarf.  Of the remaining two,
the quiescent counterpart  of one was spatially extended  and the other
counterpart coincided with  a bright elliptical galaxy).  \cite{bwb+04}
derive detection  rates of 14,  10 and $<5$ deg$^{-2}$  day$^{-1}$ for
transients  on  timescales  of  about  1900\,s in  the  $B$,  $V$  and
$R$-band, respectively.

A byproduct of the 2.2-m MPG/ESO survey for orphan afterglow was a
search for fast transients over a small
piece of sky  (exposure of 0.24\,deg$^{2}$ days;
\citealt{rgs06}).
From this survey we derive
a 3-$\sigma$ upper limit of 28 events
deg$^{-2}$ day$^{-1}$ for any class of fast transients in
the $R$ band (excluding
flares from M dwarfs), consistent with the DLS rate for fast transients.

The two DLS fast transients, if real, constitute a new class of
extragalactic transients.
The focus of this {\em Letter} is the investigation of 
fast optical transients.
Here we report on imaging and spectroscopic observations of
the two quiescent counterparts to the two DLS fast transients.

\section{The Deep Lens Survey Transient Search}
\label{sec:DLS}

The Deep Lens Survey was motivated by considerations of weak lensing.
During the period 1999-2005 five 
2$^\circ\times 2^\circ$ fields were repeatedly
observed on the CTIO-Blanco \&\ NOAO-Mayal 4-m telescopes
with exposures from 600\,s to 900\,s in the $B$, $V$, $R$
and $z'$-bands.  The repeated visits to
these fields made the survey attractive to identify fast transients.

\cite{bwb+04} reported the  analysis of a transient search  based on 14
runs conducted during the period 2000 November and 2003 April. As with
other transient searches these authors found
that the  largest number of transients were moving
solar  system objects,  followed   by  variable  stars  and  finally,
supernovae  (SNe).   

Three fast transients remained after filtering out the above
mentioned contaminants.
OT\,20020115
was detected as a bright
source, $B$=20.73\,mag, with  a turn-on time  of less than
700\,s.    It  faded   rapidly  following   a  power   law   index  of
1.5$\le \alpha \le$2.4.   Spectroscopy  established  this  to  be  a
Galactic dM4 star, presumably a UV  Ceti type flare star at a distance
of $\sim$350\,pc.

OT\,20010326 rose from obscurity to  $B$=22.7 in less than 700\,s.  It
faded on  a timescale of  1000\,s with $0.8\le\alpha\le 1.2$.   At the
position of the  transient a compact (not resolved  in an archival HST
F606W  image) persistent object  with $V\sim  24.5$, $R\sim  23.3$ and
$z'\sim 21.2$  was found. This  source resides (in projection)  on the
outskirts  of the elliptical  galaxy PKS\,1358$-$11,  a member  of the
galaxy  cluster A\,1836  at $z=0.037$.   If OT\,20010326  does  lie in
A\,1836  then the  quiescent counterpart  could be  a  bright globular
cluster associated  with PKS\,1358$-$11 or a faint  galaxian member of
A\,1836.  Radio observations were  conducted four days later resulting
in a limit of $-0.1\pm 0.3\,$mJy in the 8.5-GHz band.

OT\,20030305 rose in less than 700\,s and then doubled in flux in less
than 700\,s to $V=21\,$mag.  The transient
dropped by  1.5\,mag over the next  700\,s, and was  found to brighten
again in  the $B$-band by  0.5\,mag where it remained  nearly constant
for at least 1000\,s.  A post-cursor was detected at $R$$\sim$24.6 and
$z' \sim 21.4\,$mag. \cite{bwb+04}  claimed an apparent  ellipticity of the
post-cursor.  If so, it is  tempting to associate the quiescent object
to be a distant galaxy.

If the extragalactic nature  of OT\,20010326 and OT\,20030305 could be
established then  astronomers have uncovered a new  class of explosive
transients  with   a  staggering  combined  annual   all-sky  rate  of
$R_{\mathcal  F}\sim  10^8\,{\rm  yr}^{-1}$.   The red  color  of  the
putative host galaxies  would mean that the progenitors  belong to the
old population (cf.  novae, short  hard bursts, supernovae of type Ia)
as  opposed  to   long  duration  GRBs  which  arise   only  in  young
populations.

\section {Why are DLS Fast Transients Interesting?}
\label{sec:Interesting}

An  annual  all-sky  rate  of  fast transients,  $R_{\mathcal  F}$  of
$10^8\,{\rm yr}^{-1}$ means three events  per second.  The best way to
appreciate  the potential  importance  of the  DLS  candidates is  the
comparison  of  this  rate  with  the local  ($z=0$)  rates  of  known
transients (see Table~\ref{tab:rates}). 

\begin{table}[h]
\begin{center}
\caption{Local (z=0) rates of known transients.}
\begin{tabular}{lcl}
\hline
Type        & Rate    & Ref. \\
                 & [Gpc$^{-3}$ yr$^{-1}$] & \\
\hline\hline
long soft GRBs &  $\sim 30$ & \cite{gpw+05}\\
core-collapse SNe & $\sim 5\times 10^4$ & \cite{cet99}\\
short hard GRBs & $10-10^5$  & \cite{ngf05}\\
novae & $\sim 10^8$ & see below\\
\hline
\end{tabular}
\end{center}
\label{tab:rates}
\end{table}

The observed  rate of DLS fast  transients exceeds the  $z=0$ rates if
GRBs  (short and  long) and  core-collapse  SNe by  several orders  of
magnitude.  Allowing  for larger distances,  one must appeal  e.g., to
core collapse SNe  out to a redshift of $\sim  1$.  However, the short
duration  of OT\,20010326  and OT\,20030305  make little  sense  for a
supernova origin. The DLS transients are also unlikely associated with
long duration GRBs.  The $z$-integrated rate of these  events is $\sim
10^5\,{\rm yr}^{-1}$.

We  can also  compare the  DLS fast  transient rate  to  novae.  These
explosions have the advantage of being less expensive (relative to the
above catastrophic or one-shot explosions) owing to recurrence.

At  $z=0$,  the  mean  $K_s$-band  volumetric  density  is  $5.7\times
10^{8}h\,L_{\odot(K)}\rm  Mpc^{-3}$ \citep{cnb+01};  we  set $h$,  the
ratio of the Hubble's  constant to $100\,\rm km\,s^{-1}\,Mpc^{-1}$, to
0.7. The nova  production rate expressed  in the usual units  is about
$3\times       10^{-10}\,L_{\odot(K)}^{-1}\,       {\rm      yr}^{-1}$
\citep{ws04}.  \cite{drb+94} suggest  that the  specific  incidence of
novae is  higher in star-forming galaxies.  The  local volumetric nova
rate is at least $\rho_N \sim 10^8\,{\rm Gpc^{-3}\,yr^{-1}}$ and could
be three times higher.

If the  DLS transients  were associated with  to some kind  of unusual
novae    then   the   mean    distance   would    be   $[(3R_{\mathcal
F})/(4\pi\rho_N)]^{1/3}  \sim 0.4\,$Gpc.  At  this mean  distance they
were as bright as supernovae but with rise (or fall time) much shorter
than supernovae and  even smaller than that of  the fastest novae ever
known.  In  this framework, the  DLS fast transients form  an entirely
new and truly exotic class of cosmic cataclysmic events.

Unfortunately, a  prosaic explanation is that OT\,20030305
and  OT\,20010326 are  simply  flares  from M  dwarfs  in our  Galaxy.
Motivated to  resolve this  great ambiguity we  undertook observations
with  the  Keck telescope  of  the  counterparts  of OT\,20030305  and
OT\,20010326.

\section{Keck Observations of OT\,20010326 and OT\,20030305}
\label{sec:Observations}

Our observing  program was performed  with the Low  Resolution Imaging
Spectrograph  (LRIS;  \citealt{occ+95})  mounted  on the  Keck-I  10-m
telescope and in two separate runs (see Table~\ref{tab:log}).  LRIS as
suggested by the  name, has both an imaging  and a spectroscopic mode.
The  light beam  is  split by  a  dichroic and  images/spectra can  be
obtained in the Blue and Red channels simultaneously.

\begin{table}
\begin{center}
\caption{Log of the LRIS observations.}
\begin{tabular}{lcccc}
\hline
Date             & Mode    & Setup        & Exposure     &  Seeing \\
                 &         &              & [s]          &  [arcsec]\\     
\hline\hline
{\bf OT\,20010326:}\\
2004 April 22    & Sp & 400/3400 & 2$\times$1200 & 0.9 \\
                 & Sp & 400/8500 (7830\,\AA) & 2$\times$1200 & 0.9 \\
\hline 
{\bf OT\,20030305:}\\
2006 Feb 23      & Im   & $B$          & 5$\times$330  & 1.0 \\
                 & Im   & $R$          & 4$\times$300  & 1.0 \\
2006 Feb 24      & Im   & $B$          & 5$\times$330  & 1.0 \\
                 & Im   & $B$          & 10$\times$150 & 1.0 \\
                 & Im   & $R$          & 6$\times$300  & 1.0 \\
                 & Im   & $I$          & 10$\times$150  & 1.0 \\
                 & Sp &  400/3400   & $3\times 1830$ & 1.0 \\
                 & Sp &  400/8500 (8300\,\AA) & $3\times 1800$ & 1.0 \\
\hline
\end{tabular}
\tablecomments{The first column is the UT date of the observation. The
mode (imaging  (Im) or spectroscopy  (Sp)) is indicated in  the second
column.  In column three we  list the imaging filter (if imaging mode)
or the grism/grating (number of  grooves/blaze angle) and (for the red
grating)  the central  wavelength.   Columns four  and  five give  the
exposure sequence and the full width at half maximum in the $R$-band.}
\end{center}
\label{tab:log}
\end{table}

\subsection{OT\,20010326}

\begin{figure}
\psfig{figure=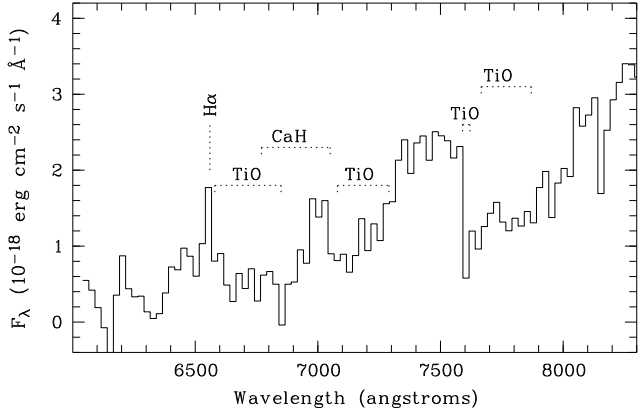,width=9.1truecm,angle=-90}
\caption{LRIS   spectrum  of  the   quiescent  counterpart  of
OT\,20010326.   The  H$\alpha$ emission  line  and CaH+TiO  absorption
bands  are  marked. The  spectrum  is  binned by  a  factor  of 25  to
emphasize the broad absorption features.}
\label{fig:OT0326}
\end{figure}

Owing to the faintness  ($R$$\sim$23.3) and red color ($V-R=1.2\,$mag)
the   blue  channel  ($\lambda<5600\,$\AA)   spectrum  is   of  little
value. The red channel spectrum shows strong absorption features which
we  identify  with TiO  and  CaH  bands (Figure~\ref{fig:OT0326}).   A
narrow emission feature close to the H$\alpha$ wavelength is detected.
We identify  this with H$\alpha$. The  line is unresolved  with a full
width at half maximum  (7$\pm$2\,\AA) consistent with the instrumental
line width.   The centroid is blue  shifted with respect  to the local
standard  rest frame  by $140\pm  20\,{\rm  km\,s}^{-1}$. The  measured
H$\alpha$    line   flux    is   $(2.3\pm    0.3)\times   10^{-17}{\rm
erg\,cm^{-2}\,s^{-1}}$.

\subsection{OT\,20030305}

\begin{figure*}
\centerline{\psfig{file=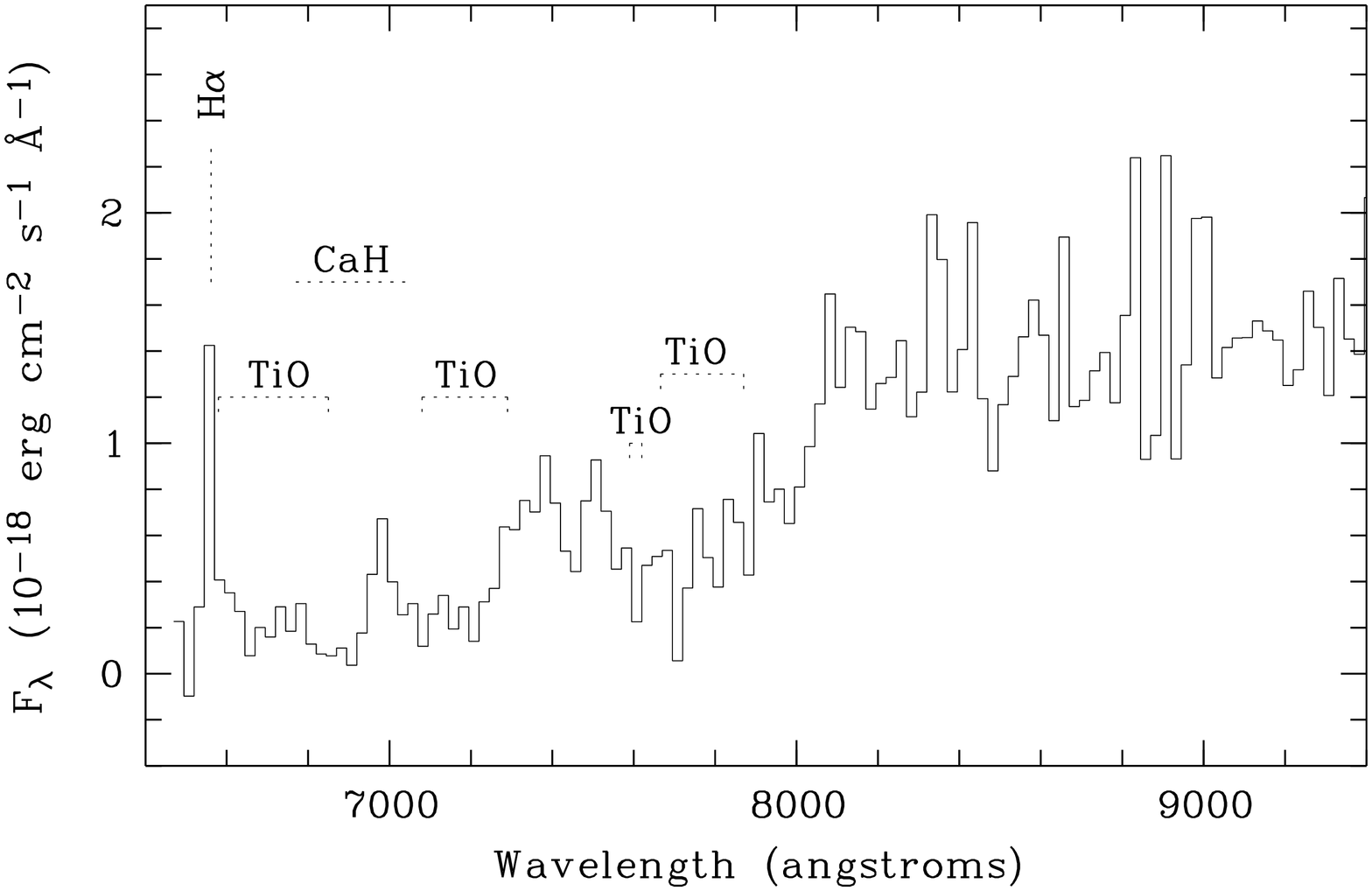,width=3.5in,angle=-90}
\psfig{file=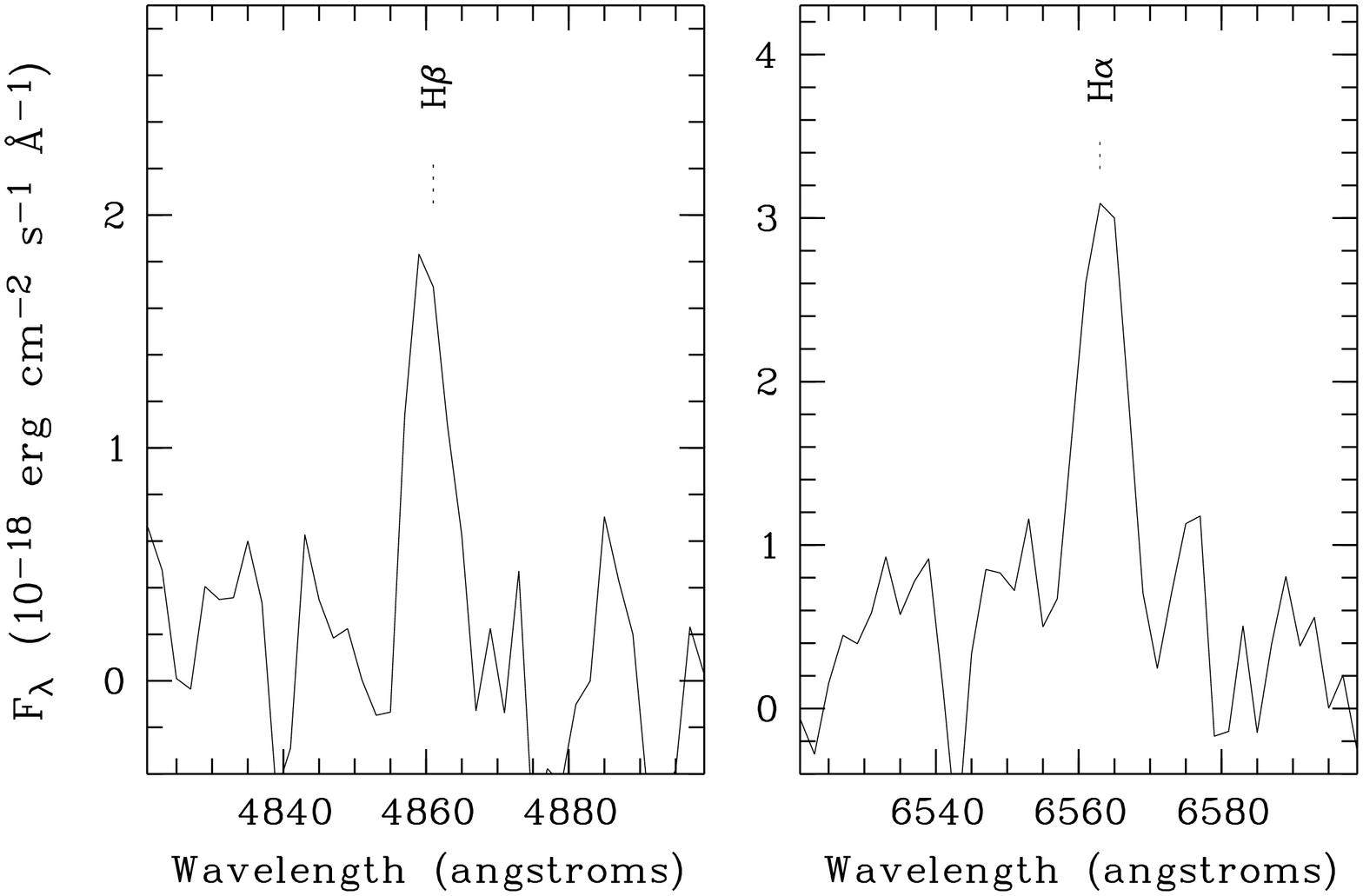,width=3.5in,angle=-90}}
\caption{LRIS   spectrum    of   the   quiescent   counterpart   of
OT\,20030305. The right panel shows the heavily binned (factor 25) red
continuum  together  with  the  H$\alpha$ emission  line  and  CaH+TiO
absorption bands marked.  Close-ups of the regions around H$\beta$ and
H$\alpha$ are shown in the center and right hand panel, respectively.}
\label{fig:OT0305}
\end{figure*}

We undertook deep observations of  this source. The red spectrum shows
a   strong  emission   feature  which   we  identify   with  H$\alpha$
(Figure~\ref{fig:OT0305}).  Fortified with  the H$\alpha$ detection we
were also  able to detect  H$\beta$ emission in  the blue part  of the
spectrum (Figure~\ref{fig:OT0305}, middle and right panels).  The full
width  at half  maximum of  the H$\alpha$  line is  $7\pm 0.8\,$\,\AA,
comparable  to   the  instrumental  resolution.    The  line  centroid
corresponds to a velocity with respect to the local standard rest frame
of  $-5\pm 10\,{\rm  km  s}^{-1}$.  The  H$\alpha$  and H$\beta$  line
fluxes   were   measured   as   $(2.6\pm   0.2)\times   10^{-17}\,{\rm
erg\,cm^{-2}\,s^{-1}}$   and    $(1.1\pm   0.2)\times   10^{-17}\,{\rm
erg\,cm^{-2}\,s^{-1}}$, respectively.

The  imaging data  were  calibrated  with respect  to  the DLS  source
catalog and verified  using the colors of stars  in the field compared
to colors derived  from the synthetic stellar spectra  included in the
library of \citet{p98}. The  quiescent counterpart was not detected in
the  $B$-band  down  to  a  limiting magnitude  of  $B=26.6\,$mag  but
detected  in   the  redder  bands  with   $R=24.54\pm  0.10\,$mag  and
$I=22.17\pm  0.07\,$mag.   As  the  seeing  was  moderate  during  our
observations   and   the  imaging   data   suffered  from   additional
instrumental distortions,  we are unable to confirm  (or disprove) the
claimed ellipticity of the source.

\section{Discussion \&\ Conclusion}
\label{sec:Discussion}

In the  Deep Lens Survey  Transient Search three fast  transients
with rise or  fall time of $\sim$1000\,s were  discovered. Surprisingly
all three were detected in the $B$-band  and none  were seen  in
the  $R$-band.  Similarly,  all three events  had  quiescent   red
counterparts.   \cite{bwb+04}  performed subsequent  spectroscopic
follow-up  observations  of  the  brightest transient  and identified
it  as a  flare  from a  Galactic M  dwarf.

Here we  reported Keck spectroscopic  and imaging observations  of the
quiescent counterparts  of the remaining  two events. The  spectrum of
OT\,20010326 shows strong  TiO absorption band heads and  is thus an M
star.   The  spectrum of  OT\,20030305  shows  H$\alpha$ and  H$\beta$
emission  lines   and  is  consistent  with  a   Galactic  origin  (no
significant local  standard rest frame velocity).   The emission lines
and the red color of the quiescent counterpart are best interpreted as
arising from a Galactic M dwarf.

Assuming a distance of about 1\,kpc for the OT\,20030305 and
OT\,20010326 (based on their apparent $R$ magnitude) the flare
energy release is $5\times 10^{33}\,$erg.  This is comparable to
the energy release from the UV-Ceti variable YY Gem \citep{m74}.
However as (reasonably) pointed by \cite{bwb+04} one expects that
high-altitude ($z$) M dwarfs are not expected to possess strong
activity (owing to their older age).  It may well be that a fraction
of late type dwarfs retain activity for a longer period.

Alternatively, the late type counterparts  discussed here may have
had activity induced later in their life. Indeed we know of several
classes of detached binaries belonging to the old Population and
which are active (e.g.,  RS CVn;  \citealt{hall76}). 
We suggest that the quiescent
DLS counterparts are rapidly spinning owing to binarity (driven by a
close-in planet or a brown dwarf companion).  Unfortunately, the
spectral resolution of our LRIS spectrum pose no significant
constraint on the rotation period: the full width at half maximum
merely constrain the rotation period to be larger than the breakup
period.

The key conclusion  of our paper is that it is 
now clear that all the 
fast transients found by
the  DLS are  flares  from  Galactic M  dwarfs. The fast transient
annual rate in $B$ band (obtained by 
using the  sky exposure and  transient detection efficiency  given in
\citealt{bwb+04}) is  $R_{\mathcal F}  \sim 10^8\,{\rm
yr}^{-1}$. 
The impressive value of
$R_{\mathcal F}$, relative to any known category of extragalactic
transients (\S\ref{sec:Interesting}), means that the 
``foreground fog'' of flares will  
ensure that the false positives outnumber
genuine extragalactic events by at least two orders of magnitude.

Indeed, current searches even with modest coverage
are already facing a steady drizzle of
interlopers. Noting the tendency of  astronomers to report
all unusual transients as candidate GRB afterglows we scoured the
GRB literature. A large flare in the localization of the
gamma-ray burst GB 7810006B turns out to be a dMe star
\citep{gm95}.
\citet{msp+00} reported a source
brightening to $R=19.5\,$mag in the IPN localization
of GRB 001212; the quiescent counterpart was  
later identified with a 2MASS M dwarf 
\citep{Gizis00}. 
During the course of the Catalina survey for near earth asteroids
\citet{Christensen04} identified 
a fast transient source (fading from $R=15.9$ to 19.7\,mag in 1 hr)
in Lynx and later spectroscopically established to be an M dwarf
\citep{dgp04}. The fast transient found in the 
CFHT Legacy Survey
\citep{mab+05} was later 
identified with a 2MASS counterpart, likely a 
M dwarf \citep{Price05}.

Consider a specific example of 
a survey undertaken  by  a telescope  with  field of  view,
$\Delta\Omega$ (in  steradians). Let $\tau$  be the exposure  time per
field of  view (frame time, usually between 30\,s and 300\,s).  The number of  flares per frame  is thus
$N_{\mathcal F}=R_{\mathcal F}\tau \Delta\Omega/(4\pi)$ and the number
over  a  time  $T$  is  $T  R_{\mathcal  F}\Delta\Omega/(4\pi)$.   For
$\Delta\Omega=2.8\times   10^{-3}\,$steradian  (corresponding   to  10
square  degrees) and $T=0.3\,$yr  (a year's  worth of  observing)  we find
$N_{\mathcal F} \sim 6,600$. 

How  can   this  immense  fog   of  M  dwarf  flares   be  penetrated?
Spectroscopy of  the quiescent counterparts  would require significant
amount  of  time  on large  telescopes  and  is  thus not  a  feasible
approach.  A counterpart which has spatial extent may simply be due to
an  asterism  or a  gaseous  nebula  formed  by repeated  flares.   An
attractive locale  such as  a bright galaxy  or a cluster  of galaxies
(cf.  OT\,20010326) does not  necessarily imply  extragalactic origin.
Finally, those transients which lie in unfavorable regions (close to a
diffraction  spike,  close  to  a  bright object)  would  have  poorer
photometry and errors  may conspire to make the  transient appear more
interesting.   Indeed, it  is  only transients  with such  interesting
coincidences  or  properties that  will  inevitably  lure the  hapless
observer into a web of incorrect conclusions.

The best line  of defense is to take advantage of the  blue (UV)
color of flares.  On   detection  of  a  potential   fast  transient
follow-up observations in  a band (or  two) complementary to the
discovery band should  be  undertaken and  the  color(s)  thus  used
to  discriminate against the  M dwarfs.  The Rayleigh-Jeans tail
of a blackbody  has a distinctly different spectrum ($f_\nu\propto
\nu^2$) as opposed to say a  synchrotron radiation  ($f_\nu\propto
\nu^\alpha$  with $\alpha\sim -1$).  Needless  to say these
observations should be done  whilst the flare is in progress i.e.
on timescales of 100\,s or so.

We close this paper with two observations. First, Pan-STARRS
and  LSST have  a  multitude  of science  goals.  Discovering new
classes of fast  transients is only one of them.  From the work of
\cite{bwb+04} and the work presented here it is now abundantly clear
that flares from late type stars constitute a veritable foreground
fog.  Surveys have to be tuned to minimize the fog:
judicious choice of
directions (high Galactic and ecliptic latitude), careful 
choice of the primary search (optical) filter, limiting the target
fields so that (over time) M dwarfs in those fields are identified,
undertaking a pre-search survey designed to identify suspect late
type dwarfs  (e.g. deep $K$ band observations during bright time)
and/or a well planned rapid followup strategy with complementary
filters may be more successful than multi-purpose surveys.

Next, we note that the OT\,{\it yyyymmdd} designation is  Y2K compliant
and  thus represents a  slight improvement over  the present GRB
naming convention. Optical astronomers have the opportunity to
continue the  (quixotic and impractical) GRB  and SN  convention
of  appending  A, B,  C, ..., AA and so on  (for successive  events)
but,  given the  measured  rate of  flares from M dwarfs of
$10^8\,{\rm yr}^{-1}$,  face the risk of drowning in a blizzard of
alphabets.

\begin{acknowledgements}
We thank A. Becker and P. Boeshaar for constructive feedback and
S. Djorgovski for reminding us of the Lynx transient which then led
us to search the GCN literature for other examples of M dwarf flares.
We acknowledge useful discussions with M.~della Valle, E.~Ofek,
E.~S.~Phinney, P.~Price, E.~Priest, M.~Salvato \&\  E.~Scannapieco.
This work is supported in part by grants from NSF and NASA.
\end{acknowledgements}

\clearpage

\clearpage

\clearpage

\clearpage


\begin{thebibliography}{21}
\expandafter\ifx\csname natexlab\endcsname\relax\def\natexlab#1{#1}\fi

\bibitem[{{Becker} {et~al.}(2004){Becker}, {Wittman}, {Boeshaar},
  {Clocchiatti}, {Dell'Antonio}, {Frail}, {Halpern}, {Margoniner}, {Norman},
  {Tyson}, \& {Schommer}}]{bwb+04}
{Becker}, A.~C., et al., 2004, Astrophys. J., 611,
  418

\bibitem[{{Cappellaro} {et~al.}(1999){Cappellaro}, {Evans}, \&
  {Turatto}}]{cet99}
{Cappellaro}, E., {Evans}, R., \& {Turatto}, M. 1999, \aap, 351, 459

\bibitem[{{Christensen}(2004)}]{Christensen04}
{Christensen}, E. 2004, {GCN Circular} 2849

\bibitem[{{Cole} {et~al.}(2001){Cole}, {Norberg}, {Baugh}, {Frenk},
  {Bland-Hawthorn}, \& {et al.}}]{cnb+01}
{Cole}, S., et al., 2001, Mon. Not. R. astr. Soc., 326, 255

\bibitem[{{della Valle} {et al.} (1994){della Valle}, {Rosino}, {Bianchini}, \& {Livio}}]{drb+94}
{della Valle}, M., {Rosino}, L., {Bianchini}, A., \& {Livio}, M., 1994, \aap, 287, 403

\bibitem[{{Djorgovski} {et~al.}(2004){Djorgovski}, {Gal-Yam}, \&
  {Price}}]{dgp04}
{Djorgovski}, S.~G., {Gal-Yam}, A., \& {Price}, P. 2004, {GCN Circular} 2851

\bibitem[{{Gizis}(2000)}]{Gizis00}
{Gizis}, J.~E. 2000, {GCN Circular} 901

\bibitem[{{Greiner} \& {Motch}(1995)}]{gm95}
{Greiner}, J. \& {Motch}, C. 1995, Astr. Astrophys., 294, 177

\bibitem[{{Groot} {et~al.}(2003){Groot}, {Vreeswijk}, {Huber}, {Everett},
  {Howell}, {Nelemans}, {van Paradijs}, {van den Heuvel}, {Augusteijn},
  {Kuulkers}, {Rutten}, \& {Storm}}]{gvh+03}
{Groot}, P.~J., et al., 2003,
  \mnras, 339, 427

\bibitem[{{Guetta} {et~al.}(2005){Guetta}, {Piran}, \& {Waxman}}]{gpw+05}
{Guetta}, D., {Piran}, T., \& {Waxman}, E. 2005, Astrophys. J., 619, 412

\bibitem[{{Hall}(1976)}]{hall76}
{Hall}, D.~S. 1976, in ASSL Vol. 60: IAU Colloq. 29: Multiple Periodic Variable
  Stars, ed. W.~S. {Fitch}, 287

\bibitem[{{Kaiser} {et~al.}(2002){Kaiser}, {Aussel}, {Burke}, {Boesgaard},
  {Chambers}, {Chun}, {Heasley}, {Hodapp}, {Hunt}, {Jedicke}, {Jewitt},
  {Kudritzki}, {Luppino}, {Maberry}, {Magnier}, {Monet}, {Onaka}, {Pickles},
  {Rhoads}, {Simon}, {Szalay}, {Szapudi}, {Tholen}, {Tonry}, {Waterson}, \&
  {Wick}}]{kab+02}
{Kaiser}, N., et al., 2002, in Survey and Other Telescope
  Technologies and Discoveries. Edited by Tyson, J. Anthony; Wolff, Sidney.
  Proceedings of the SPIE, Volume 4836, pp. 154-164 (2002)., ed. J.~A. {Tyson}
  \& S.~{Wolff}, 154--164

\bibitem[{{Malacrino} {et~al.}(2005){Malacrino}, {Atteia}, {Boer}, {Klotz},
  {Veillet}, {Cuillandre}, \& {Kaavelars}}]{mab+05}
{Malacrino}, F., {Atteia}, J., {Boer}, M., {Klotz}, A., {Veillet}, C.,
  {Cuillandre}, J., \& {Kaavelars}, J. 2005, {GCN Circular} 2964

\bibitem[{{Moffett}(1974)}]{m74}
{Moffett}, T.~J. 1974, Astrophys. J. Supp. Series, 29, 1

\bibitem[{{Mohan} {et~al.}(2000){Mohan}, {Sagar}, {Pandey}, \&
  {Castro-Tirado}}]{msp+00}
{Mohan}, V., {Sagar}, R., {Pandey}, S.~B., \& {Castro-Tirado}, A.~J. 2000, {GCN
  Circular} 900

\bibitem[{{Nakar} {et~al.}(2005){Nakar}, {Gal-Yam}, \& {Fox}}]{ngf05}
{Nakar}, E., {Gal-Yam}, A., \& {Fox}, D.~B. 2005, astro-ph/0511254

\bibitem[{{Oke} {et~al.}(1995){Oke}, {Cohen}, {Carr}, {Cromer}, {Dingizian},
  {Harris}, {Labrecque}, {Lucinio}, {Schaal}, {Epps}, \& {Miller}}]{occ+95}
{Oke}, J.~B., et al., 1995, \pasp, 107, 375

\bibitem[{{Pickles}(1998)}]{p98}
{Pickles}, A.~J. 1998, \pasp, 110, 863

\bibitem[{{Price}(2005)}]{Price05}
{Price}, P.~A. 2005, {GCN Circular} 2965

\bibitem[{{Rau} {et~al.}(2006){Rau}, {Greiner}, \& {Schwarz}}]{rgs06}
{Rau}, A., {Greiner}, J., \& {Schwarz}, R. 2006, Astr. Astrophys., 449, 79

\bibitem[{{Williams} \& {Shafter}(2004)}]{ws04}
{Williams}, S.~J. \& {Shafter}, A.~W. 2004, Astrophys. J., 612, 867

\end{thebibliography}
\end{document}